\documentclass[final,5p,times,twocolumn]{elsarticle}

\usepackage{amssymb}
\usepackage[utf8]{inputenc}
\usepackage[T1]{fontenc}
\usepackage{lmodern}
\usepackage{booktabs}
\usepackage{textcomp}
\usepackage{graphicx}
\usepackage{subfigure}
\usepackage{xspace}
\usepackage{nicefrac}
\usepackage[greek,english]{babel}
\usepackage{textgreek}
\usepackage{hyperref}
\usepackage{amsmath}
\usepackage{flushend}

\journal{Journal of Microscopy}

\begin{document}

\begin{frontmatter}

\title{A python tool to determine the thickness of the hydrate layer around clinker grains using SEM-BSE images.}

\author[1]{Florian Kleiner\corref{corauthor}}
\author[2]{Franz Becker}
\author[1]{Christiane Rößler}
\author[1]{Horst-Michael Ludwig}
\cortext[corauthor]{Corresponding author}
\address[1]{F. A. Finger-Institute for Building Materials Science, Bauhaus-University Weimar, Germany}
\address[2]{Faculty of Science - Chair of Mineralogy, University of Erlangen-Nuernberg (FAU), Germany}

\begin{abstract}
To accurately simulate the hydration process of cementitious materials, understanding the growth rate of C-S-H layers around clinker grains is crucial.
Nonetheless, the thickness of the hydrate layer shows substantial variation around individual grains, depending on their surrounding.
Consequently, it is not feasible to measure hydrate layers manually in a reliable and reproducible manner.
To address this challenge, a software has been developed to statistically determine the C-S-H thickness, requiring minimal  manual interventions for thresholding and for setting limits like particle size or circularity.

This study presents a tool, which automatically identifies suitable clinker grains and and perform statistical measurements of their hydrate layer up to a specimen age of 28 days.
The findings reveal a significant increase in the C-S-H layer, starting from 0.45\,\textmu m after 1 day and reaching 3.04\,\textmu m after 28 days.
However, for older specimens, the measurement of the C-S-H layer was not feasible due to limited pore space and clinker grains.
\end{abstract}

\begin{keyword}
hydration \sep calcium silicate hydrate \sep hydrate layer
\end{keyword}

\end{frontmatter}

\section{Introduction}\label{sec1}

Understanding the formation of the micro-structure of cementitious materials allows to model and thus improve their properties.
Calcium-silicate-hydrate (C-S-H) and portlandite (CH) are the main hydration products of ordinary Portland cement.
C-S-H is mainly responsible for the concrete strength. 
Hence, its growth rate within the first 28 days is of major interest to model the hydration process \cite{NguyenTuan.2022,NguyenTuan.2020}.

The growth process can be observed in embedded and polished binder samples by using a scanning electron microscope (SEM) in backscatter electron (BSE) mode (see Figure \ref{fig1}).
Analyzing the phase distribution by segmenting each identifiable phase is a common practice \cite{Chu.2021}. 
Furthermore, it is fairly easy to measure the particle size distribution of the alite (impure form of tricalcium silicate, C$_3$S) particles \cite{Lyu.2019}.

During the hydration of cementitious materials, C-S-H forms a layer around clinker grains, and its thickness increases over time.
This layer can be differentiated into ($i$) the dense, inner product and ($ii$) the less dense, needle like shaped outer product \cite{Garrault.2006}.
Especially the development of the outer product is often limited by the surrounding particles or by other developing phases like C-S-H or portlandite (CH).
Bazzoni \cite{Bazzoni.2014} demonstrated that is is possible to determine the C-S-H needle length using images from a scanning transmission electron microscope (STEM). 
Nevertheless, it is challenging to measure the hydrate layer thickness (HLT) manually in a reliable and reproducible manner.
Hence, in this study, we established a workflow and software tool to automatically determine the HLT.

\section{Materials and Methods}\label{sec2}

In this study, a commercially available alite (MIII polymorph, Vustah, Czech Republic) was utilized.
The chemical composition (Table \ref{tab:xrf}) was determined by X-ray fluorescence spectroscopy (XRF), and the phase purity was confirmed at 99.4 ± 0.4 wt.-\% (alite) using X-ray diffraction (XRD) analysis.

\begin{table}[]
\caption{Chemical composition of the utilized alite obtained by XRF. \label{tab:xrf}}
\begin{tabular}{@{}lr@{}}
\toprule
Oxides  & Alite in wt.-\%   \\ \midrule
SiO$_2$          & 26.34 \\
TiO$_2$          & 0.031 \\
Al$_2$O$_3$      & 0.238 \\
Fe$_2$O$_3$      & 0.094 \\
Mn$_2$O$_3$      & 0.006 \\
MgO              & 1.700  \\
CaO              & 70.79 \\
Na$_2$O         & 0.083 \\ \bottomrule
\end{tabular}
\end{table}

Alite pastes were prepared with a water-to-solid ratio of 0.5 and stored in sealed cylindrical containers with a diameter of approx. 8\,mm.
The phase composition and alite purity of similarly prepared samples stored for 1, 7, and 28 days were assessed by quantitative XRD analysis using the G-factor method \cite{Jansen.2011}.
The measurements were carried out using a Bruker D8 DaVinci diffractometer (MA, USA).

After 1, 7, 14, 28, 84, and 365 days the hydration of the cylindrical prisms was stopped by immersion in isopropanol and subsequent drying at 60$^\circ$C.
The specimens were cut, embedded in low-viscosity resin and mechanically polished using an oil-based diamond paste with a grain size down to 0.25\,\textmu m.
Finally, the polished specimens were carbon coated (approx. 10\,nm) to avoid charging effects in the SEM.
The specimens were transferred to a SEM (Helios G4 UX, ThermoFisher Scientific, MA, USA) to capture high-resolution BSE images of their surfaces using the built-in concentric backscatter detector (CBS).
These image sets were stitched to larger images.
Afterwards, they were processed using custom code, which is the main subject of this study.
The code was developed for Python 3.11 and was integrated into a Jupyter Notebook and it is publicly available \cite{Kleiner.2023c}.

\begin{figure}[htb!]
\centerline{\includegraphics[width=\columnwidth]{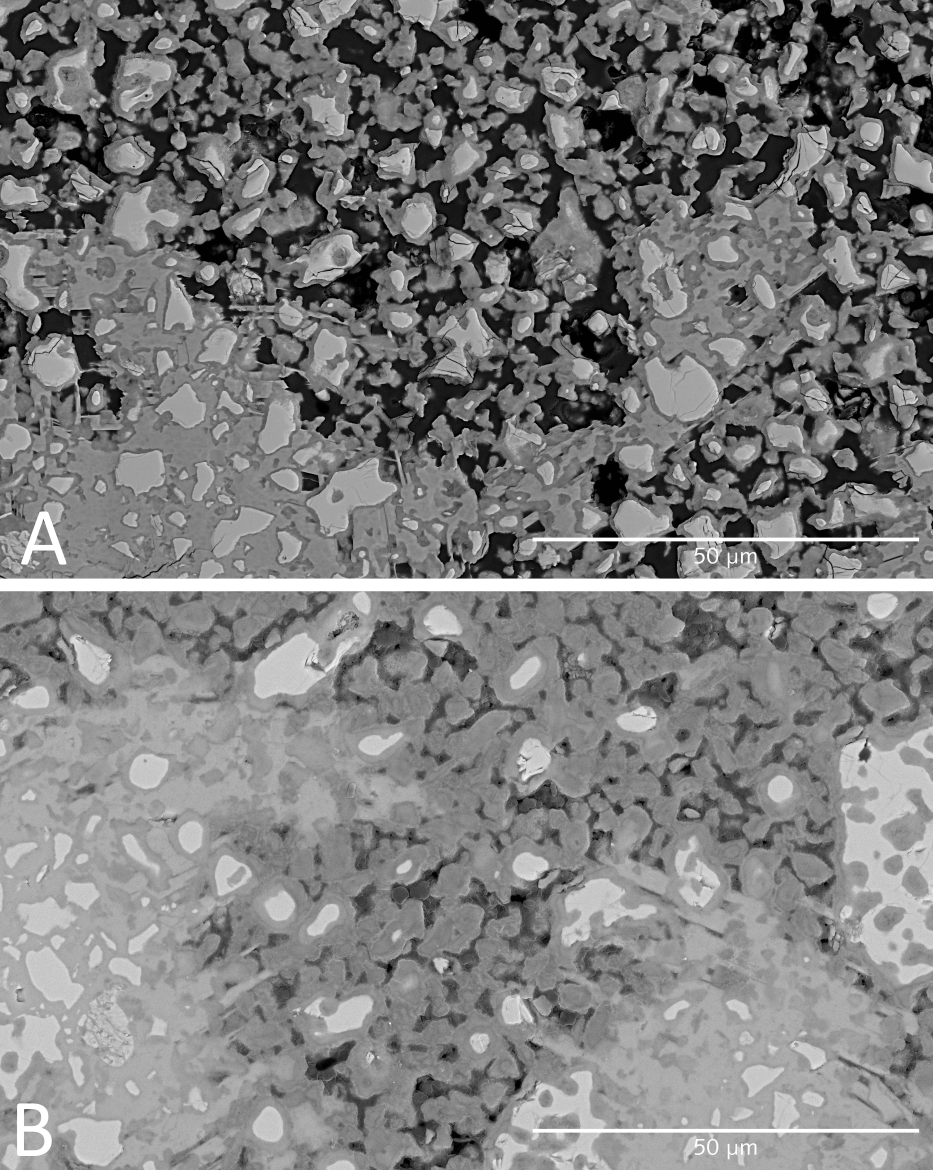}}
\caption{Detail of the stitched, unprocessed BSE images of embedded and polished alite sections after 7 (A) and 28 days (B) of hydration. Identifiable phases from dark to light grey: resin/pores, C-S-H, CH, alite. The images also show a typical deviation in image contrast, even after using identical imaging parameters (12\,kV, 0.8\,nA). \label{fig1}}
\end{figure}

The image contrast was logarithmically enhanced using the library \textit{scikit-image} \cite{vanderWalt.2014}.
Subsequently, the images were denoised employing the Non Local Means algorithm \cite{Buades.2011} implemented in the library \textit{OpenCV} \cite{Bradski.2000}.
The raw and preprocessed images have varying contrasts as shown in Figure \ref{fig1}A and B.
Even after preprocessing, the segmentation cannot be carried out using fixed threshold values. 
Therefore, the images were segmented using two manually selected threshold values into three phases: ($i$) pores, ($ii$) hydration products (CH and C-S-H), and ($iii$) unhydrated clinker.
The hydration products were not separated since this is not always possible without major segmentation errors.
However, the software is also able to process pre-segmented images (e.g. segmented by machine learning algorithms), if data is provided in a supported format (3-channel TIFF, one colour channel per phase).
Finally, the holes within particles were removed to avoid any errors caused by pores within clinker grains.

For the herein presented data, the segmented images were used to identify clinker particles with a grain diameter between 0.3 and 9.0\,\textmu m (grain area between 0.07 to 63.6\,\textmu m$^{2}$).
Furthermore, particles with a circularity lower than 0.3 were excluded to avoid extreme sectioning effects.
For very oval particles, the hydrate fringe could otherwise be greatly overestimated and undercuts could also lead to undesirable results.

In addition, the clinker particles shrink over time and small grains are consumed.
Since the selected grain diameter was kept constant regardless of the specimen age, this effect is not taken into account in this analysis.

\begin{figure*}[tb]
\centerline{\includegraphics[width=\textwidth]{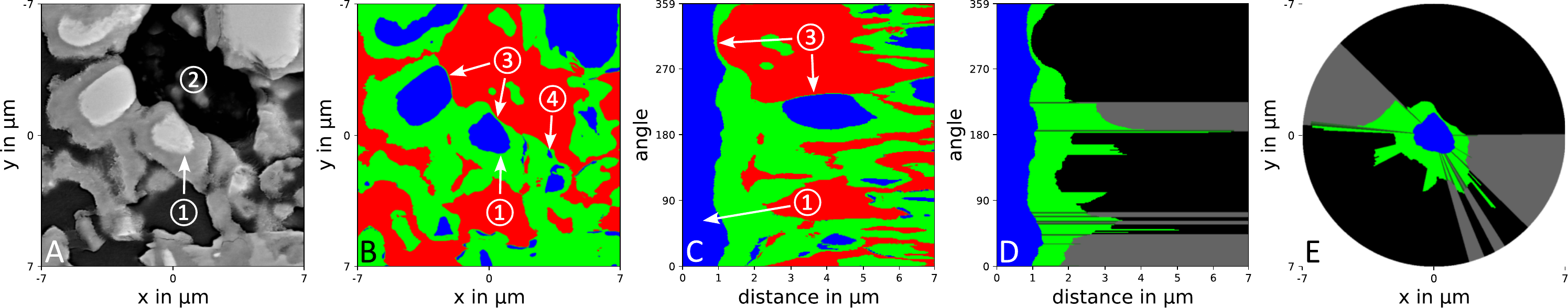}}
\caption{Illustration of the processing that is done to measure the HLT. In the center of A, B and E is an unhydrated clinker (1) grain within alite hydrated for 7 days. A: raw image; B: segmented image (red: pores, green: hydrates, blue: alite); C: transformation of B into polar coordinates, the clinker grain (blue) is on the left; D: like C, but only with the identified hydrate in green (grey areas were excluded from processing); E: version of D transformed into Cartesian coordinates.\label{fig2}}
\end{figure*}

For each particle, a circular region with a radius of 7.0\,\textmu m was selected (Figure \ref{fig2}A, B).
This image was then transformed into polar coordinates (Figure \ref{fig2}C).
Particles located too close to the image border (less than 7\,\textmu m) were excluded from further processing.
The HLT was determined using 360 steps, representing a 1$^\circ$ angle per step.
To achieve this, the distance from the alite phase (left-hand side of Figure \ref{fig2}C, blue) to the next pore (red) was measured.
If there was no adjacent pore within the 7.0\,\textmu m radius, the measurement was stopped (Figure \ref{fig2}D, areas with adjacent grey).
To reduce errors, 3$^\circ$ before and after those areas were ignored as well (Figure \ref{fig2}D, marked as darker vertical lines).

Subsequently, the histogram of the HLT distribution was analyzed.
To safely determine the maxima $l_\text{max}$, a function was fit to the HLT histogram using equation \eqref{eqn:fitcurve}, where $l$ is the hydrate layer thickness and $f$ is the frequency.

\begin{equation}
	\label{eqn:fitcurve}
	f = a \cdot e^{-0.5 \left( \frac{log(l-d)-b}{c} \right)}
\end{equation}

Additionally, by using the area of the alite $A_{\text{C}_3\text{S}}$ and the hydrate phases $A_\text{hydrate}$, obtained from the image segmentation, it was possible to estimate the amount of hydrate phase $\sigma_\text{hyd}$ (CH + C-S-H) using Equation \eqref{eqn:est_frac}.
The densities used to calculate these values were 3.14\,g/cm$^3$ ($\rho_{\text{C}_3\text{S}}$), 2.2\,g/cm$^3$ (CH) and $\approx$ 2.0\,g/cm$^3$ (C-S-H) \cite{Garrault.2006,Jennings.2000}.
For the hydrate mixture, a volume ratio C-S-H:CH of 2.5:1 was assumed, giving an approximate hydrate density $\rho_\text{hyd}$ of 2.07\,g/cm$^3$.

\begin{equation}
	\label{eqn:est_frac}
	\sigma_\text{hydrate} = \frac{A_\text{hydrate} \cdot \rho_\text{hydrate}}{A_\text{hydrate} \cdot \rho_\text{hydrate} + A_{\text{C}_3\text{S}} \cdot \rho_{\text{C}_3\text{S}}}
\end{equation}

\section{Results and Discussion}\label{sec4}

The herein presented method was applied to alite specimens of different ages in order to test its applicability.
However, an existing dataset (1 day) and datasets explicitly created for this research (7, 14, 28, 84, and 365 days) were mixed.
The datasets differ mainly in the imaged area and slightly in SEM imaging parameters.
For example, the accelerating voltage used varied from 10 to 15\,kV.
An overview of the images used is shown in Table \ref{tab:c3s}.
The raw images are publicly available \cite{Kleiner.2023b}

\begin{table}[]
\footnotesize
\caption{Basic information and selected results of the alite datasets processed. Image size, normed particle count (diameter <\,9.0\,\textmu m, circularity >\,0.3), phase areas based on the segmentation, fit parameters ($a$-$d$), and the maximum HLT $l_\text{max}$ of the second value accumulation. \label{tab:c3s}}
\begin{tabular*}{\linewidth}{@{}lrrrrrr@{}}
\toprule
age in days				& 1		& 7		& 14		& 28		& 84		& 365 \\ \midrule
area in mm$^{2}$		& 0.56	& 2.41	& 2.41	& 2.41	& 2.41	& 2.41 \\
particles\\per mm$^{2}$	& 25296	& 6540	& 17719	& 8050	& 8385	& 7512 \\ \midrule
pores in\\area-\%		& 38.5	& 28.5	& 22.2	& 13.7	& 21.7	& 14.8 \\
hydrates in\\area-\%	& 38.0	& 52.8	& 44.5	& 78.7	& 72.7	& 83.2 \\
clinker in\\area-\%		& 23.9	& 19.1	& 33.6	&  7.7	& 5.7	& 2.0 \\ \midrule
$a$ in \textmu m		&  0.999	&  0.929	& 0.898	&  0.898	& -		& -  \\
$b$						& $-$1.506	& 0.097	& 0.585	&  0.858	& -		& -  \\
$c$						&  1.003	&  0.471	& 0.487	&  0.475	& -		& -  \\
$d$						&  0.219	&  0.297	& 0.680	& $-$1.000 & -		& - \\ \midrule
$l_\text{max}$ in \textmu m	&  0.45	&  1.43	& 1.69	&  3.04	& -		& - \\ \bottomrule
\end{tabular*}
\end{table}

On close inspection of the images, it becomes evident that the low-viscosity resin did not completely penetrate the specimens.
This is apparent in Figure \ref{fig2}A (2), where the large pore on the top center appears a little darker than the surface of the epoxy resin in other parts of the image.
In the most cases, this is not an issue, as those holes are interpreted as pores anyway.
However, larger unfilled pores should be manually painted black, since this could lead to erroneously identified clinker particles or hydrates.
Additionally, the needle-like structure of the C-S-H and the penetration depth of the electron beam (from 1.60\,\textmu m in alite up to 3.75\,\textmu m in epoxy resin at 12 kV \cite{kleiner.2021}) cause a brightness gradient between pores and hydrate phase, particularly in denoised images.

\begin{figure*}[htbp]
\centerline{\includegraphics[width=\textwidth]{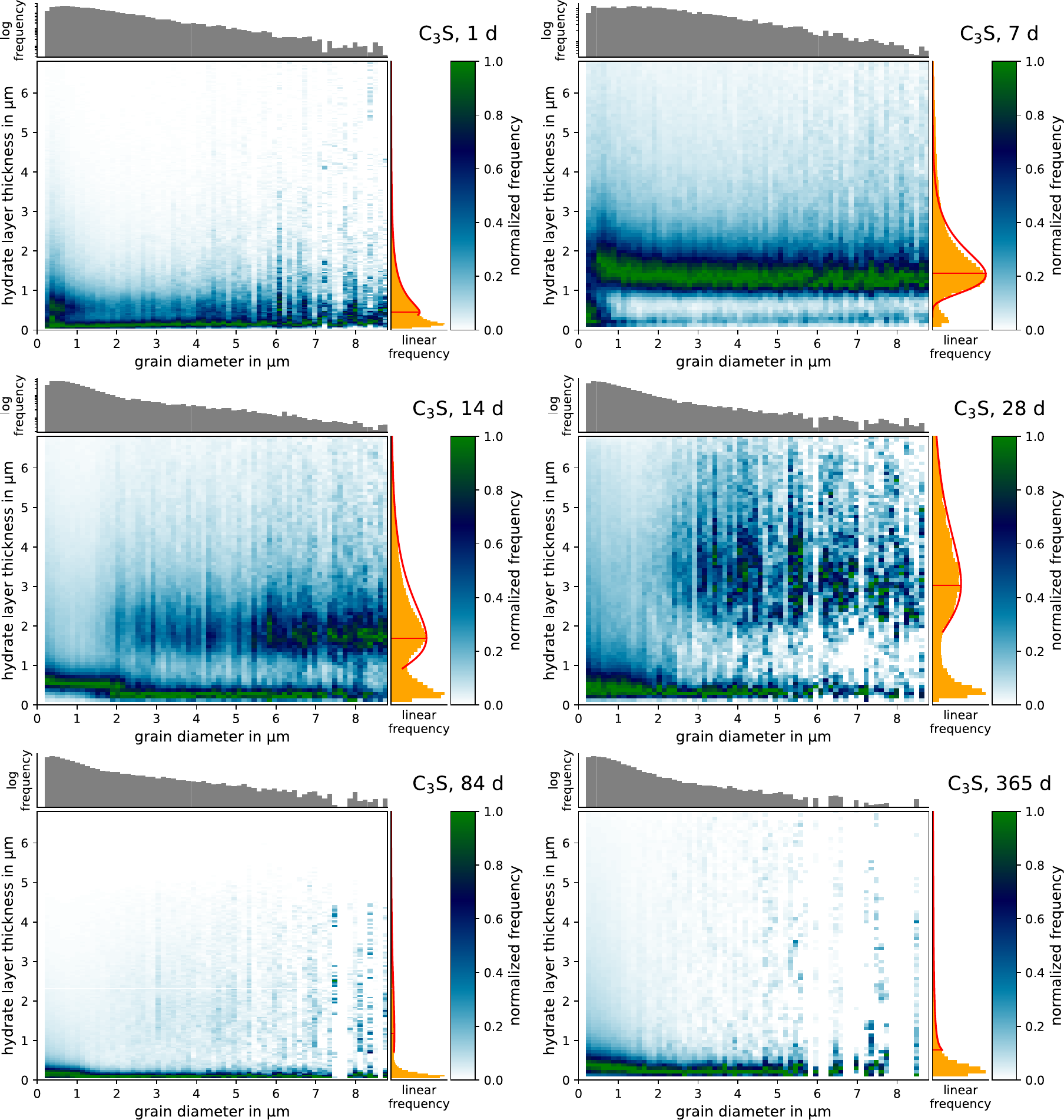}}
\caption{Normalized distributions of the HLT of hydrated alite (aged 1, 7, 14, 28, 84, and 365\,days) depending on the particle diameter.
The histogram on top illustrates the logarithmic frequency for each diameter bin in the diagram below.
The histogram on the right illustrates the frequency of HLT measurements as sum of the main diagram.
The values are thus independent of the frequency of single grain diameters.\label{fig:results2d}}
\end{figure*}

A significant challenge lies in accurately determining the threshold value that distinguishes the pore space from the hydrate phase (see (3) in Figure \ref{fig2}B and C).
The polishing process leaves the specimen surface with slight unevenness due to differences in material hardness, resulting in undesirable topography and brightness variations at phase borders. 
This segmentation artifact, combined with cracks within the clinker particles might cause the first peak of the HLT measurements (see orange distributions in Figure \ref{fig:results2d}).
Figure \ref{fig2}B (4) also indicates that in some instances, edges of hydrate were incorrectly identified as clinker phase, which may also affect the later analysis.
These issues are further enhanced, because the datasets were stitched together from several smaller images, whose brightness may show slight deviations.

If the HLT is compared to the grain diameters (Figure \ref{fig:results2d}), it is noticeable that the data density decreases with increasing age for the same examined area size.
This is indicated by an increasing level of noise, particularly in the case of larger grain diameters.
This phenomenon can be attributed to the decrease in the amount of unhydrated particles present in older specimens (see Table \ref{tab:c3s}), leading to shifts in the particle size distribution over time.
The grey histogram above the 2D-distribution, which shows the logarithmic frequency of measurements per grain diameter bin, also indicates this alteration.

These 2D-distributions indicate accumulations of HLT measurements, depending on the alite grain diameters.
To eliminate the afore mentioned effect of the particle size distribution, the the maximum amount of hydrate layer measurements was normalized to 1 for each particle diameter step.
These normalized distributions were then used to calculate the cumulative HLT distribution depicted in orange on the right of the 2D-distributions in Figure \ref{fig:results2d}.

All distributions show an accumulation at very low measurements (< 0.5 to 1.0\,\textmu m), which could be an artifact caused by the preparation and segmentation.
For example, a very small hydrate fringe appears to be around some particles after the automatic segmentation (compare Figure \ref{fig2}B c, (3)).
However, this layer would not exist after a manual segmentation.

Histograms of specimens with an age up to 28 days show two clusters of HLTs.
However, only the second accumulation seems to be a valid result.
Therefore, the first accumulation was ignored.
For older specimens this can be observed only for larger particles (e.g. a diameter above  2.0\,\textmu m for 14 and 28 days hydrated alite), while smaller particles only seem to have the first accumulation.
Nevertheless, a trend of an increasing HLT between 1 and 28 days can be observed.

To obtain the value of $l_\text{max}$, Equation \eqref{eqn:fitcurve} was fitted to the orange HLT histogram in Figure \ref{fig:results2d}.
The resulting curves are plotted in red.
However, due to the absence of a second accumulation in specimens older than 28 days, fitting was not feasible in those cases.
The resulting fit parameters $a$, $b$, $c$, $d$, and $l_\text{max}$ are listed in Table \ref{tab:c3s}.

The selected limit value for circularity of 0.3 was a compromise between a high data density for larger particles and signal clarity.
Increasing this value can improve the signal clarity of the second accumulation, if there are enough suitable particles available.
Furthermore, this reduces the chance of overestimating the HLT due to angular distortions for non-circular particles.
However, this also reduces the amount of measures, especially for particles with a diameter larger than 5.0\,\textmu m.
This is an issue for older specimens, where the particle count is significantly reduced due to the hydration progress.
For example, in order to be able to make any statement at all about specimens hydrated for 84 and 365 days, the circularity value must be lowered to at least 0.4 to obtain enough data for particles larger than 5.0\,\textmu m in diameter.
Nevertheless, they do not seem to have a second cluster of HLTs in the range measured.

\begin{figure}[htb]
\centerline{\includegraphics[width=\columnwidth]{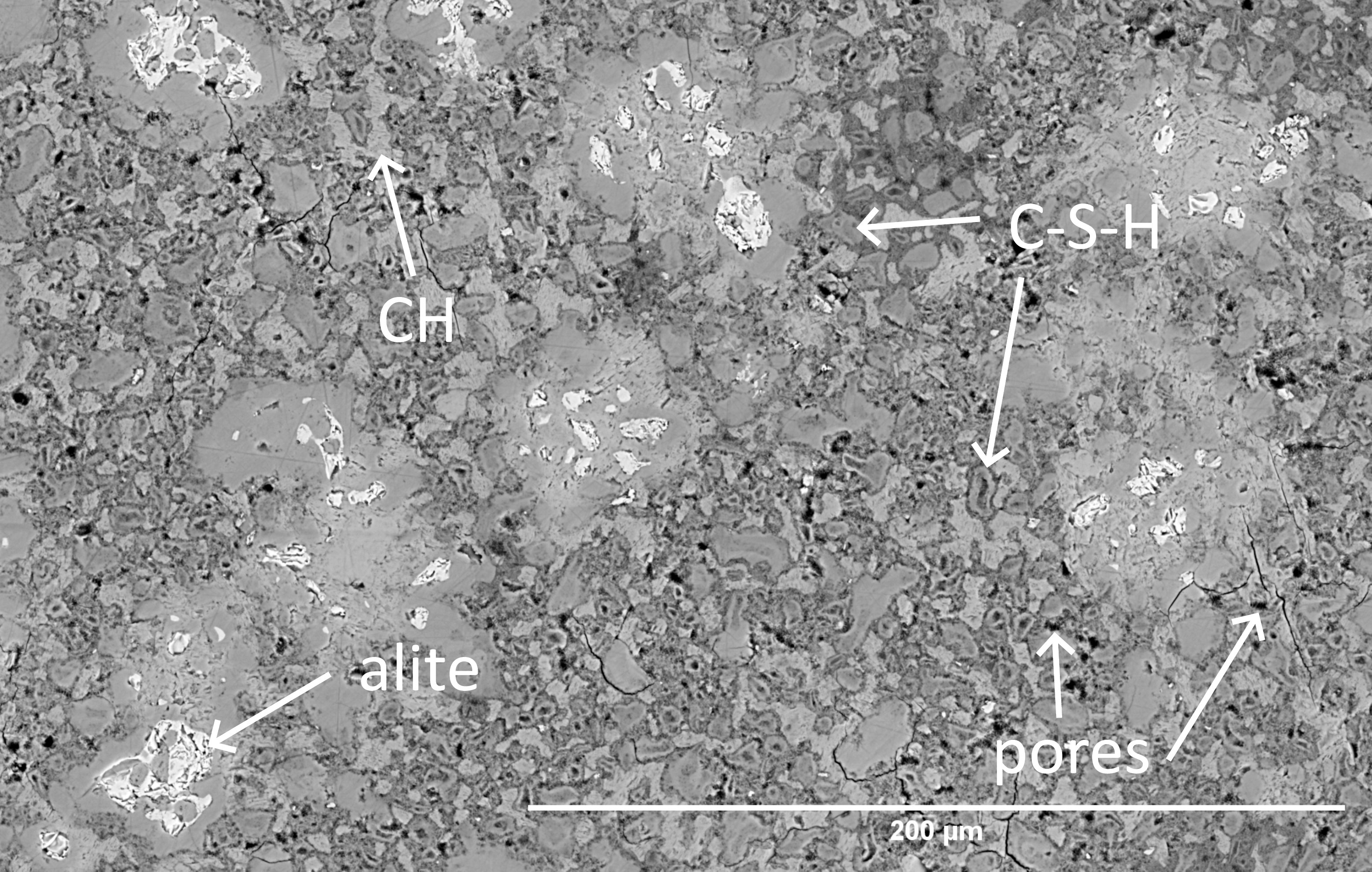}}
\caption{Image detail of 365 days hydrated alite. This image shows a lack of pore space filled by CH, disintegrated larger particles and significantly smaller particles than in Figure \ref{fig2} (note the scale bar differences). \label{fig:1year}}
\end{figure}

There are multiple possible reasons for this result, which can be followed on the basis of Figure \ref{fig:1year}.
Firstly, the pore space between hydrates is significantly smaller, reducing the possibility to obtain good measurements.
Secondly, the few remaining large particles appear to disintegrate into several smaller particles or break up due to sample preparation, leaving some apparent pores.
Finally, a large part of the remaining space is filled by CH.
Therefore, in this case the results should be considered invalid.

\begin{figure}[htb]
\centerline{\includegraphics[width=\columnwidth]{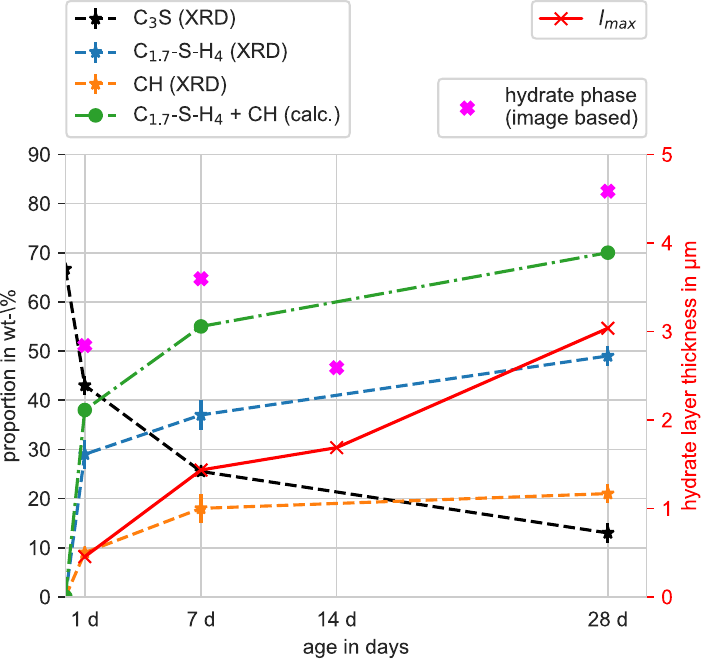}}
\caption{Comparisson between the hydration progression measured using XRD (dashed lines) with $l_\text{max}$ (solid line).
The pink markers (X) indicate the amount of hydrate phase $\sigma_\text{hyd}$ estimated from the image segmentation.\label{fig:comp_hydration}}
\end{figure}

It is important not to confuse the increase of the HLT with the hydration progress.
Figure \ref{fig:comp_hydration} shows the hydration progress, measured using XRD (dashed lines).
The combined ratio of C-S-H and CH is represented by the green dash-dotted line, the HLT is plotted in a solid red line.
While both, the XRD and HLT measurements indicate an increase, the HLT does not follow the same pattern as the XRD measurements.

This can be argued with multiple reasons:
\begin{itemize}
\item First and foremost, completely dissolved grains are not taken into account to measure the hydrate layer thickness.
This concerns a significant area of the image and becomes more prevalent with specimen age.
\item The grains could be cut at an unfavourable position (far off-center), which leads to an apparently larger hydrate layer.
\item CH and C-S-H phases are not separated. However, this is not anticipated to pose a significant problem, as the majority of measurements will be automatically excluded due to the absence or minimal presence of pores within the 7.0\,\textmu m radius, particularly when the particle is embedded in CH.
\end{itemize}

In comparison, Masoero et al. \cite{Masoero.2014} calculated a reaction zone of about 0.4 to 0.5\,\textmu m.
They used different particle size distributions of alite in a suspension (water/solid ratio of 50).
In their proposed model, the reaction zone should be filled after 21\,h.
This seems to be in the same range of the measurement of about 0.45\,\textmu m in this study.

In contrast, for sieved C$_3$S (mean particle size 6\,\textmu m) Costoya Fernández \cite{CostoyaFernandez.2008} calculated a peak thickness of C-S-H of 0.3\,\textmu m after 12\,h of hydration.
Powders with larger particle size distributions showed a significantly smaller C-S-H thickness (C-S-H thickness of 0.1\,\textmu m for a mean particle size of 18\,\textmu m).

Garrault et al. \cite{Garrault.2006} proposed a maximum dense C-S-H-layer thickness of 0.4\,\textmu m, with a tendency to even lower values for larger grain diameters.
Bazzoni \cite{Bazzoni.2014} measured the C-S-H needles on top of the dense C-S-H and found, they were in the range of 0.3 to 0.4\,\textmu m in length after 7 days.
Combining both results, the overall HLT should therefore not exceed 0.8\,\textmu m, which is not supported by the data found in this study.

In some studies \cite{Masoero.2014, CostoyaFernandez.2008, Garrault.2006}, the C-S-H layer thickness was calculated from ion concentrations and heat flow measurements.
They are therefore based on volumetric measurements of the whole specimens and not individual alite grains.
In contrast, our study only includes C-S-H layers directly around a particle and ignores a bulk of the formed hydrate.

The results of the estimated hydrate phase based on image segmentation are shown in Figure \ref{fig:comp_hydration} (pink markers (X)).
They do not give a clear pattern, but usually the amount of hydrate was overestimated.
This could be due to the fact that $\sigma_\text{hydrate}$ was calculated from areas $A_\text{hydrate}$ and $A_{\text{C}_3\text{S}}$ instead of their respective volumes and since only a rough phase ratio and density estimate was used.
However, this can not explain the deviation in the data for the 14 day specimen.
Therefore, it is likely that not enough area was measured for a reliable phase measurement.

Finally, the computational requirements and processing time of the script used are non-negligible factors.
The images with an area of 2.4\,\textmu m$^2$ were scaled from a pixel size of 0.0422\,\textmu m to 0.0843\,\textmu m to reduce processing time and memory requirements. 
The processing of the resized image requires about 8\,GB of RAM and can take up to 12 hours on an Intel \textit{Core i7 8700}, depending on the image size and the amount of particles to be processed.
However, there is still some optimization potential by improving the multi-threaded processing or by compiling the code to machine language.

\section{Conclusions and outlook}\label{sec5}

The herein presented method can be utilized to determine the development of the HLT of alite for young to medium aged specimens (up to 28 days) in an objective manner.
Aside from the image segmentation itself, it is a fully automated process, requiring little user intervention.

The C-S-H growth with increasing age is clearly reflected in the results.
The maxima in the HLT distribution increases from 0.45\,\textmu m after 1 day up and reaching 3.04\,\textmu m after 28 days.
However, at 84 days or later the HLT distribution does not allow to evaluate the HLT distribution using the selected parameters.
Consequently, the proposed approach finds applicability when the microstructure is not overly dense. In practical terms, this signifies the method's viability requires the presence of a substantial amount of pores and clinker particles for its successful application.

Additionally, it was tried to calculate the hydration progress from the generated data.
However, BSE images up to 2.4\,mm$^2$ in size still seem to be too small for a reliable evaluation.

While the script was only applied to a simplified cementitious system (clean alite). 
First tests not presented in this paper did indicate that this method may also be applicable for clean belite specimens.
Furthermore, it should be expandable to more complex multiphase materials.
However, this imposes higher demands on the segmentation quality.
Applying more advanced segmentation methods (e.g. machine learning based) could therefore improve the reliability of the results.

In the future, the influence of parameters like the circularity limit should be studied using synthetic datasets.
Furthermore, multiple images of specimens of varying sizes but the same age should be studied, to determine the specimen-to-specimen variability.
Finally, this method could be improved to determine the HLT of 3-dimensional, voxel-based datasets.

\section*{Author contributions}

Florian Kleiner: Conceptualization, Methodology, Investigation, Software, Visualization, Data curation, Writing – original draft.
Franz Becker: Supporting measurements (XRF, XRD), Writing – review \& editing, Validation
Christiane Rößler: Supervision, Writing – review \& editing, Validation.
Horst-Michael Ludwig: Financing, Supervision, Writing – review \& editing, Validation.

\section*{Acknowledgment}

The research was supported by the Deutsche Forschungsgemeinschaft (DFG), grant number 344069666.
Open access funding enabled and organized by project DEAL.
Thanks to M. Böhme, S. Unbehau and F. Ellermann for their helpful input.

\section*{Conflict of interest}

The authors declare no potential conflict of interests.

\section*{Supporting information}

Additional supporting information may be found in the online version of the article at the publisher’s website.

\bibliographystyle{elsarticle-num} 
\bibliography{arXiv}

\end{document}